\pdfoutput=1

\documentclass[11pt,twoside,a4paper,cmspaper,final,collab]{cms-tdr}
\def\svnVersion{171946}\def\cmsCernNo{2012-364}\def\cmsCernDate{\today}\def\cmsMessage{Submitted to Physics Letters B}

\begin{document}\cmsNoteHeader{EXO-12-015}
\typeout{**** Begin including local definitions.tex file *****}
\typeout{**** tdr loads this quietly without a specific input command *****}

\newcommand{\fix}[1]{{\bf <<< #1 !!!} }

\newcommand{\statUnc}[1]{\ensuremath{\pm{#1}(\mathrm{stat.})}}
\newcommand{\systUnc}[1]{\ensuremath{\pm{#1}(\mathrm{syst.})}}
 
\newcommand{\Zzero}{\ensuremath{{\rm Z}}}
\newcommand{\mumu}{\ensuremath{\mu^+ \mu^-}}
\newcommand{\tautau}{\ensuremath{\tau^+ \tau^-}}
\newcommand{\Mmumu}{M_{\mu\mu}}
\newcommand{\DYto}[1]{\ensuremath{{\rm DY} \rightarrow #1}}
\newcommand{\zPrime}{\ensuremath{{\rm Z^\prime}} }
\newcommand{\zpmumu}{\ensuremath{\zPrime\rightarrow\mu\mu}}
\newcommand{\zPrimePsi}{\ensuremath{Z{}^\prime_{\psi} }}
\newcommand{\zPrimeSSM}{\ensuremath{{\mathrm
Z}^\prime_{\textrm{\sc{ssm}}}} }
\newcommand{\zPrimeStSM}{\ensuremath{{\mathrm Z}^\prime_{\textrm{\scriptsize{S\lowercase{t}}}}} }

\newcommand{\gkk}{\ensuremath{G_\mathtt{KK}}}

\newcommand\ipb{{{\rm pb}}^{-1}}
\newcommand\ifb{{{\rm fb}}^{-1}}

\newcommand{\anaLumi}{\ensuremath{4.1~\ifb}}
\newcommand{\anaLumiAllGood}{\ensuremath{3.6~\ifb}}
\newcommand{\collisionEnergy}{\ensuremath{ \sqrt{s} = 8\TeV }}
\newcommand{\swRelease}{CMSSW $5\_2\_3$}

\newcommand{\mcGlobalTag}  {START52\_V9}
\newcommand{\dataGlobalTag}{GR_R_52_V7}

\newcommand{\MPlanck}{\ensuremath{ \overline{M}_{\rm Pl} }}

\typeout{**** End including local definitons.tex file *****}

\hyphenation{had-ron-i-za-tion}
\hyphenation{cal-or-i-me-ter}
\hyphenation{de-vices}

\RCS$Revision: 162634 $
\RCS$HeadURL: svn+ssh://svn.cern.ch/reps/tdr2/papers/EXO-12-015/trunk/EXO-12-015.tex $
\RCS$Id: EXO-12-015.tex 162634 2012-12-26 13:17:31Z alverson $
\newlength\cmsFigWidth
\ifthenelse{\boolean{cms@external}}{\setlength\cmsFigWidth{0.85\columnwidth}}{\setlength\cmsFigWidth{0.4\textwidth}}
\ifthenelse{\boolean{cms@external}}{\providecommand{\cmsLeft}{top}}{\providecommand{\cmsLeft}{left}}
\ifthenelse{\boolean{cms@external}}{\providecommand{\cmsRight}{bottom}}{\providecommand{\cmsRight}{right}}
\providecommand{\ZPSSM}{\ensuremath{\cPZpr_\mathrm{SSM}}}
\providecommand{\ZPPSI}{\ensuremath{\cPZpr_\psi}}
\cmsNoteHeader{EXO-12-015}
\title{Search for heavy narrow dilepton resonances in pp collisions at $\sqrt{s}$ = 7\TeV and $\sqrt{s}$ = 8\TeV}

\date{\today}

\abstract{
An updated search for heavy narrow resonances decaying to muon or
electron pairs using the CMS detector is presented.
Data samples from pp collisions at $\sqrt{s}$ = 7\TeV and 8\TeV
at the LHC,
with integrated
luminosities of up to 5.3
and 4.1\fbinv, respectively, are combined.
No evidence for a heavy narrow resonance is observed.
The analysis of the combined data sets excludes, at 95\% confidence level,
a Sequential Standard Model $\cPZpr_\mathrm{SSM}$ resonance lighter than 2590\GeV, a superstring-inspired
$\cPZpr_{\psi}$ lighter than 2260\GeV, and Kaluza--Klein gravitons lighter than 2390
(2030)\GeV, assuming that the coupling parameter $k/\overline{M}_\text{Pl}$ is 0.10 (0.05).
These are the most stringent limits to date.
}

\hypersetup{%
pdfauthor={CMS Collaboration},%
pdftitle={Search for heavy narrow dilepton resonances in pp collisions at sqrt(s) = 7 TeV and sqrt(s) = 8 TeV},
pdfsubject={CMS},%
pdfkeywords={CMS, physics, resonances, Z', muons, electrons}}

\maketitle

\section{Introduction}

A number of scenarios for physics beyond the standard model
predict the existence of
heavy narrow resonances
that decay to lepton pairs.
In this Letter, we report on
a search for resonances with the
Compact Muon Solenoid (CMS) detector at the CERN Large Hadron Collider
(LHC)~\cite{lhc}.
We consider the following three benchmark scenarios:
the Sequential Standard Model $\ZPSSM$ with
standard-model-like couplings~\cite{Altar:1989}, the $\ZPPSI$ predicted by grand unified
theories~\cite{Leike:1998wr}, and Kaluza--Klein graviton excitations in the Randall--Sundrum
(RS) model of extra dimensions~\cite{Randall:1999vf, Randall:1999ee}.
The RS model has two free parameters.
One parameter is  the mass of the first graviton excitation, and the other is the coupling
$k/\MPlanck$, where $k$ is the curvature of the extra dimension and $\MPlanck$
is the reduced
Planck scale.

Previous searches for narrow $\cPZpr \rightarrow \ell^+ \ell^- (\ell=\mu,\Pe) $ resonances
have been reported by the CMS~\cite{CMS-dilep-2011} and
ATLAS~\cite{ATLAS-dilep-2011-5fb} Collaborations, each based on integrated luminosities of 5\fbinv
at $\sqrt{s} = 7$\TeV.
The CDF and D0 experiments have published results based on integrated luminosities exceeding
5\fbinv of $\Pp\Pap$ collisions at
$\sqrt{s}=$ 1.96\TeV~\cite{CDF_Zp,CDF_RS,D0_RS,D0_Zp,CDF_RSele,CDF_SSM}.
The best previous direct lower limits on the $\ZPSSM$ and $\ZPPSI$ masses
are 2330\GeV and
2000\GeV~\cite{CMS-dilep-2011}, respectively.
The best previous direct limits on RS graviton ($G_\mathrm{KK}$) production are 2160\GeV for
$k/\MPlanck = 0.1$~\cite{ATLAS-dilep-2011-5fb} and 1810\GeV for
$k/\MPlanck = 0.05$~\cite{CMS-dilep-2011}.
Indirect constraints~\cite{opal,delphi,l3,aleph} are less stringent.

We use data samples from pp collisions at center-of-mass energy $\sqrt{s} = 8$\TeV,
corresponding to an integrated
luminosity of $3.6\pm0.2\fbinv$ for the dielectron channel.
The dimuon channel does not use information from the calorimeters and
incorporates additional data from running periods
when the calorimeters were not fully operational.
This increases the integrated luminosity of the sample to $4.1\pm0.2$\fbinv.
We combine the analysis of these data with previous results from the analysis based on an integrated luminosity of 5\fbinv at $\sqrt{s} = 7$\TeV~\cite{CMS-dilep-2011}.
The reconstruction, selection criteria, efficiencies, and systematics for the two
data sets are very similar.
The results are applicable to any model with a narrow resonance
that has equal dimuon and dielectron branching fractions.
A resonance is considered narrow if the experimental width is dominated by the detector resolution.

We perform a shape-based analysis of the dilepton mass spectra searching for a peak
on a smoothly falling distribution with the overall background normalization determined
by an unbinned maximum likelihood fit.
The data are consistent with
expectations from the standard model.
We report limits on the ratio  ($R_{\sigma}$) of the production cross sections
times branching fractions of a heavy narrow resonance to that of the
$\Zzero$ boson, at the 95\% confidence level (CL).
Many experimental and theoretical uncertainties cancel in this ratio.
We further translate these limits into lower limits on the masses of new
heavy narrow resonances, using next-to-next-leading-order (NNLO) cross
section calculations \cite{Gavin:2010az} for the Z boson production.

\section{The CMS detector}
\label{sec:detector}

A detailed description of the CMS detector can be found in Ref.~\cite{:2008zzk}.
We briefly discuss the systems most relevant to this analysis.
The central feature of the CMS detector is an all-silicon inner tracker system,
composed of silicon pixel and strip detectors.
The tracker is surrounded by a lead-tungstate crystal electromagnetic calorimeter (ECAL) and
a brass-scintillator hadron
calorimeter (HCAL).
The finely segmented ECAL consists of nearly 76000 lead-tungstate crystals
which provide coverage up to pseudorapidity $|\eta|=3.0$.
It is divided in the barrel ($|\eta|<1.479$) and endcap ($1.479<|\eta|<3.0$)
detectors.
We define pseudorapidity as $\eta = -\ln \left[ \tan \left( \theta/2 \right) \right]$.
Here, $\theta$ is the polar angle with respect to the direction of the
counterclockwise proton beam.
We use $\phi$ for the azimuthal angle of a track's momentum
at the point of closest approach to the beamline.
The tracker and calorimeter systems reside within a 6\unit{m} diameter superconducting solenoid, which produces a 3.8\unit{T} axial magnetic field.
Muons are detected by gas-ionization chambers embedded in the steel flux-return yoke.

The CMS experiment utilizes a two-level trigger system.
The first level of the trigger (L1)
selects events of
interest using custom hardware processors~\cite{L1TDR}.
It uses information from the muon and calorimeter systems to reduce the readout
rate from the 20\unit{MHz} bunch crossing rate to a maximum rate of 100\unit{kHz}.
The software based high-level trigger (HLT) further reduces the recorded event
rate to a few hundred Hz by adding information from the inner tracker and
analyzing event information in greater detail~\cite{HLTTDR}.
\section{Event selection and object reconstruction}

The event selection
closely mirrors the one used for the $\sqrt{s} = 7$\TeV analysis.
We briefly review the procedure here.
Dimuon events are triggered by requiring at least one muon to be reconstructed
by the HLT and to have transverse momentum $\pt > 40\GeV$ and $|\eta| < 2.1$.
Dielectron events
are accepted by a double-electron trigger requiring
two clusters in the ECAL, each with transverse energy $\ET > 33\GeV$.
The trigger allows only small deposits of energy
in the HCAL
to be associated with the ECAL clusters.
HLT clusters are required to be loosely matched to the trajectories of tracks having hits in the pixel detector.
The lepton trigger efficiencies are measured using a
``tag-and-probe'' technique at the $\Zzero$ resonance~\cite{EWK-10-002-PAS,MUO-10-004-PAS,
CMS-dilep-2011}, up to transverse momenta of roughly 500\GeV for muons and
100\GeV for electrons.
For higher transverse momenta, the electron efficiency is measured using
a simple trigger which requires only an ECAL cluster with $\ET>300\GeV$ to
directly monitor the trigger efficiency for selected high mass events.
In order to have a consistent trigger between the low-mass control region and the high-mass
signal region, the simple ECAL trigger is used only to validate the efficiency of the primary trigger.
The muon trigger efficiency is 97\% for events with both muons within the
trigger acceptance, across the entire range of dimuon invariant masses of
interest.
The efficiency of the electron trigger for dielectron candidates passing
the analysis selection requirements increases from 80\% at electron
$\ET=35$\GeV to a 99\% plateau at $\ET>37$\GeV.
The efficiency threshold curve of the trigger is measured using data
collected by a lower threshold trigger that is applied to approximately
every $5^{\text{th}}$ event passing the L1 part of the trigger.
Because the threshold behavior is well-determined, the offline $\ET$ selection
cut can be placed at 35\GeV, which improves the normalization to the Z peak.
Both muon and electron trigger efficiencies are within 1--2\%
of those found in  Ref.~\cite{CMS-dilep-2011}.
Standard CMS algorithms \cite{MUO-10-004-PAS,CMS-PAS-EGM-10-004,CMS-dilep-2011}
are used to reconstruct and select muon and electron candidates.
Muon candidates are formed by matching tracks in the silicon tracker to tracks
in the muon systems.
Muon tracks are required to have hits in nine or more layers of the tracker and
include at least one hit from each of the pixel and muon systems.

Muon candidates are required to be isolated in a cone about the muon direction
of $\Delta R <0.3$ in the tracker and to have $\pt > 45$\GeV and $|\eta|<2.4$.
The quantity $\Delta R$ is defined as $\Delta R = \sqrt{(\Delta \eta)^2 + (\Delta \phi)^2}$,
where $\Delta \phi$ is in radians.
The combined fit of the muon trajectory through the tracker and muon systems
provides a reliable measurement of muon momenta
extending to order 1\TeV \cite{MUO-10-004-PAS,Khachatryan:2010mw}.
Electron candidates are formed by matching ECAL clusters to reconstructed tracks,
and are required to have $|\eta| < 1.442$ and $1.560 <
|\eta| < 2.5$ for the ECAL barrel and ECAL endcap regions, respectively.
As in Ref. \cite{CMS-dilep-2011}, electrons are additionally required to have
little associated activity in the HCAL,
have a shower shape consistent with that of an
electromagnetic object, and be isolated in a cone about the electron direction
of $\Delta R <0.3$ in the calorimeter and tracker.
To account for the contamination
which varies with the number of additional proton-proton interactions
per event, the calorimeter isolation of the electron is
corrected for the average
energy density in the event \cite{rhoCorrection}.

The ECAL is capable of measuring energies in a single crystal of up to
approximately 1.7\TeV in the barrel region and 2.8\TeV in the endcap region.
Saturation effects become
an issue for resonance masses around 4\TeV, which are beyond the reach of the
current
analysis, within the framework of the models considered here.
Occasionally, anomalously large signals are observed in the barrel ECAL due
to the direct deposition of energy into the avalanche photo-diodes (APDs)
by particles transiting the detector~\cite{Petyt2012293}.
Since the APDs are normally used to detect the scintillation light produced by
the crystals, the equivalent energies of these signals can reach into the
TeV range.
The deposits are generally in a single channel and have different pulse timings
from scintillation signals.
They are rejected by cutting on the timing of the pulse and the amount of energy
recorded in neighboring crystals.

More than 99\% of the anomalous clusters are rejected with a negligible loss of
real electrons~\cite{Petyt2012293}.
When combined with the requirement of a compatible track in the silicon
tracker, this requirement reduces background from anomalous clusters to a
negligible level.
However, the energy deposit pattern of all selected high energy electrons is scrutinized
even further.
There is no evidence that any of the remaining events is due to these anomalous signals.

Previously established techniques \cite{EWK-10-002-PAS,MUO-10-004-PAS,CMS-dilep-2011}
are used to measure event reconstruction and selection efficiencies.
The efficiencies for reconstructing and selecting leptons are roughly 90\%,
for both muons and electrons with $\pt>100$\GeV.
The efficiencies include the isolation requirements and are measured with
total uncertainties of a few percent.
Part of the systematic uncertainty
cancels in the ratio of high-mass dilepton cross section to the Z boson cross section.
Studies with Monte Carlo simulated samples predict both lepton reconstruction and selection efficiencies to be
constant within 1\% for
transverse momenta $\pt \geq 100\GeV$.

Each dilepton candidate is required to have two isolated leptons of the same
flavor that pass the identification criteria described above.
When multiple dilepton candidates are present, the most energetic pair in the bunch crossing is selected.
For dimuon events, one muon must have $|\eta|<2.1$ to satisfy the trigger requirements.
For dielectron events, at least one electron must have $|\eta| < 1.442$.
This removes events
in which both electrons are in the endcap, a topology where little signal
is expected but
which has a significant background arising from misidentified jets.
Muons are required to have opposite charge, since a charge mis-assignment implies a
large mismeasurement of momentum.
The energy estimate for electrons is dominated by electromagnetic calorimeter
information and is not sensitive the momentum mis-measurement indicated by a
charge mis-assignment.
The charge requirement would also result in a few percent efficiency loss
in a region with little background and would degrade the sensitivity of the
analysis.
Therefore, we do not impose a charge requirement for dielectron candidates
at high mass.
Muon candidates are additionally required to originate from the same vertex.
The $\chi^2$ per degree of freedom for the fit to a common vertex is required to be less than 10.
The tracker-measured transverse impact parameter with respect to the beam spot
must be less than 2 mm for each muon.
The mass resolution of a dielectron candidate is predicted by Monte Carlo simulation to be approximately 1.8\% for masses above 800\GeV.
The dimuon mass resolution is 5\% (9\%) at 1\TeV (2\TeV) and increases linearly with dimuon mass.

The opening angle of the muon pair is required to be less than $\pi-0.02$ radians.
This requirement greatly reduces the cosmic ray background associated with muons
traversing the detector.

\section{Backgrounds}

The dilepton background in pp collisions at $\sqrt{s} = 8$\TeV is very
similar to that found in $\sqrt{s} = 7$\TeV collisions~\cite{CMS-dilep-2011}
even though there were significantly more interactions per bunch crossing at the higher energy.
The effect of this "event pileup" is included in our simulations of background processes and
our data-driven estimates of the background from misidentified jets.
The dominant standard model background is due to Drell--Yan production.
The shape of this contribution is determined from Monte Carlo simulation using the
\PYTHIA~v6.4~\cite{Sjostrand:2006za} event generator.
The background contribution
is normalized to the event count at the $\Zzero$ peak by counting
same-flavor dilepton candidates within the mass window $60 < m_{\ell \ell} < 120$\GeV.
The next largest
background contribution is due to other standard model processes that
produce isolated dileptons.
We consider the lepton flavor symmetric processes of $\ttbar$, $\mathrm{tW}$, $\cPZ\rightarrow \tau^{+}\tau^{-}$,
and diboson (WW, WZ, and ZZ) production when estimating this background component.
The absolute normalization and shape for these backgrounds is taken directly from Monte Carlo
simulation generated using \MADGRAPH~5~\cite{Alwall:2011uj}, \POWHEG~\cite{Nason:2004rx,Frixione:2007vw,Alioli:2010xd,Re:2010bp}, and \PYTHIA.
We validate this background prediction by comparing the ${\mathrm e}\mu$ dilepton mass spectra for
data and simulation.

Track-based lepton isolation strongly suppresses backgrounds from jets misreconstructed as leptons.
This background is almost negligible for the dimuon channel but is a significant
portion of the non Drell--Yan background in the dielectron channel.
Since misidentification of jets as leptons is more likely to occur for electrons than for muons,
electrons have additional isolation and jet discrimination requirements.
In the dielectron channel, the main contributing processes apart from Drell--Yan are dijet, W+jet, and photon+jet
production
(where the photon is misidentified as an electron).
The probability that a jet is misidentified as an electron is measured
in bins of $\ET$ and $\eta$, using a jet-dominated sample.
This probability is then used to weight events in which one electron
satisfies all selection criteria
and the other is a candidate for being a misidentified jet, to obtain the jet background prediction.
The dimuon resonance search is
susceptible to backgrounds from cosmic ray muons.
The
expected cosmic ray background for dimuons with $m_{\mu\mu} > 200$\GeV is determined from two
complementary samples.
For events in the first sample, the requirement on the dimuon opening angle is removed.
In the second sample, the impact parameter requirement on the muon tracks is not applied.
From the populations of these two samples, the remaining cosmic ray background contamination
is estimated to be less than 0.2 events.
\section{Results}

The dilepton mass distributions for events passing all the selection criteria are
shown in Fig.~\ref{fig:massSpectra}. The ``jets'' distribution illustrates the contribution of events in which at
least one jet is misreconstructed as a lepton.
This distribution is derived from data while all other components are derived from simulation.
The relative fractions of the different background components are fixed to the ratios
of their theoretical cross sections.
The total simulated background is normalized to data at the $\Zzero$ peak
($60 < m_{\ell \ell} < 120$\GeV).
The expected yields in the control region ($120  < m_{\ell \ell} < 200$\GeV)
and in the search region ($m_{\ell \ell} > 200$\GeV) are compared with observed yields
in Table \ref{tab:yields}.
The observed mass spectra and event counts agree with standard model predictions
both in shape and normalization.

\begin{figure}[htb]
\begin{center}
\includegraphics[width=0.49\textwidth]{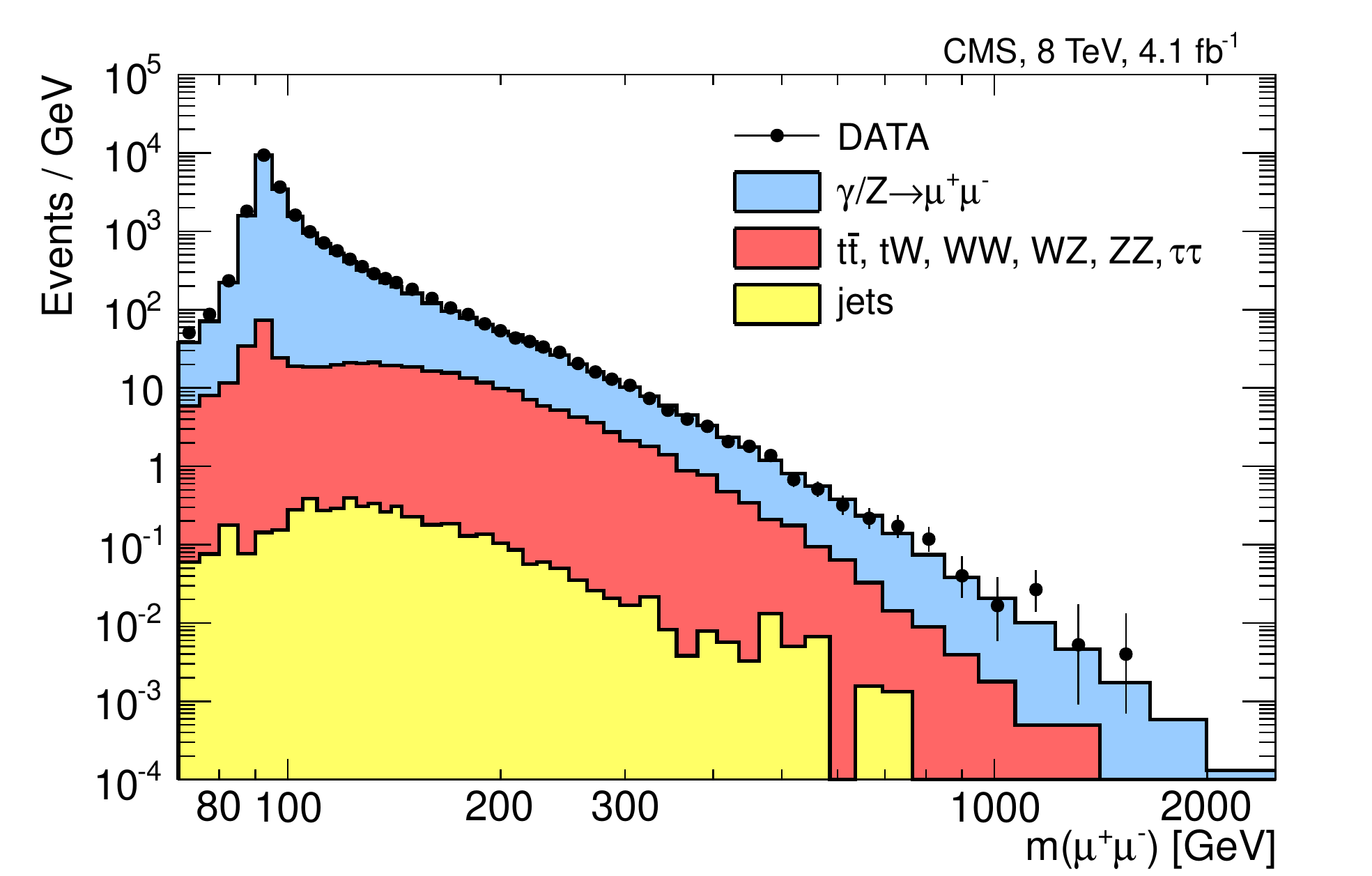}
\includegraphics[width=0.49\textwidth]{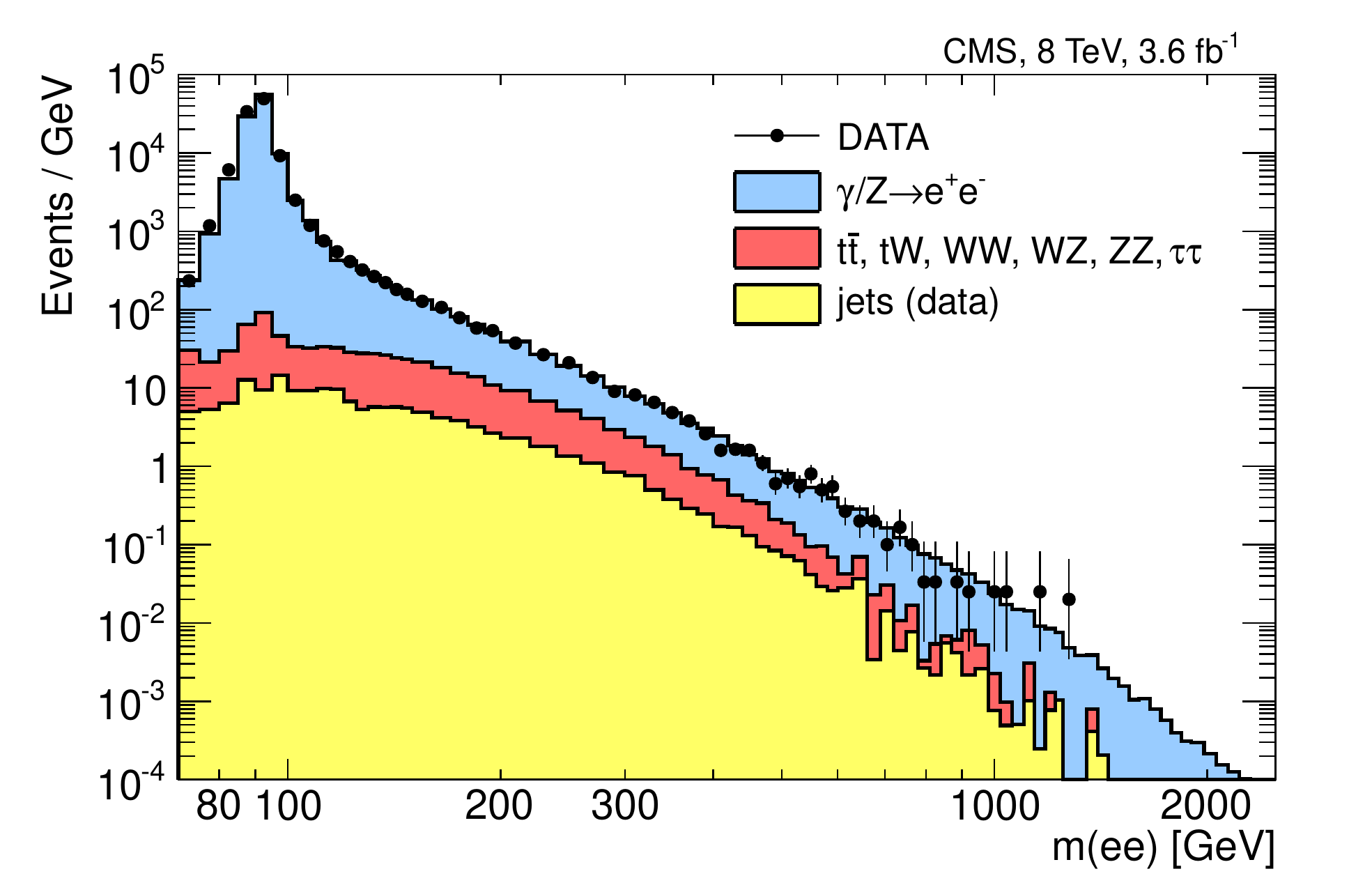}
\end{center}
\caption{\label{fig:massSpectra}
The invariant mass spectrum of $\Pgmp\Pgmm$ (left) and $\Pe\Pe$ (right) events for
the $\sqrt{s} = 8$\TeV data set. The points with error bars represent the data.
The solid histograms represent the standard model predicted background
contributions.
}
\end{figure}

\begin{table*}[!htb]
\centering
\topcaption{
The dilepton event count in the control region $120 < m_{\ell\ell} < 200$\GeV
and in the search region $m_{\ell\ell} > 200$\GeV\ for the $\sqrt{s} = 8$\TeV data set.
The total background is the sum of the events for the standard model processes listed.
Uncertainties represent a quadrature sum of statistical and systematic uncertainties.
}
\label{tab:yields}
\begin{tabular}{l|D{,}{\,\pm\,}{5.3}|D{,}{\,\pm\,}{4.3}|D{,}{\,\pm\,}{5.3}|D{,}{\,\pm\,}{4.3}}
\hline\hline
Source & \multicolumn{4}{c}{Number of events} \\
       & \multicolumn{2}{c|}{Dimuon sample }
       & \multicolumn{2}{c}{Dielectron sample} \\
 &  \multicolumn{1}{c|}{(120--200)\GeV{}} & \multicolumn{1}{c|}{$>$200\GeV{}}
 &  \multicolumn{1}{c|}{(120--200)\GeV{}}  & \multicolumn{1}{c}{$>$200\GeV{}}  \\ \hline
Data                   & 1381       & 3503      & 12030      & 2904      \\
Total background       & 13010,590  & 3630,160  & 12240,590  & 2970,260  \\ \hline
$\cPZ{/}\gamma^*$      & 11700,570  & 2920,140  & 10660,530  & 2200,220   \\
$\ttbar$ + others      &  1280,150  &  698,78   & 1220,180   & 560,80 \\
Jets                   & 26,3       &   10,1    & 360,180    & 210,110   \\
\hline\hline
\end{tabular}
\end{table*}

We set a 95\% CL limit on the ratio $R_\sigma$ of the product of the cross
section and branching fraction for each  $\cPZpr$ boson to that of the
standard model $\Zzero$ boson.
The cross section of the $\cPZpr$ boson is calculated in a window of $\pm$40\%
about the on-shell mass of the resonance, while for the $\Zzero$ boson it is calculated
in the peak window defined above.
We follow the Bayesian procedure of Ref.~\cite{CMS-dilep-2011}, which is based on an
unbinned extended maximum likelihood analysis.
We calculate the limits using the 8\TeV data alone, as well as
from a combination of the 8\TeV and 7\TeV data sets.
Mass-dependent ratios of parton distribution functions (PDF) at $\sqrt{s} = 7 $\TeV and 8\TeV
are used as an additional input to derive
limits on $R_{\sigma}$ at 8\TeV, $R_{\sigma, 8 \mathrm{TeV}}$, that combine both data sets.
The CTEQ6.1 LO PDF set~\cite{Pumplin:2002vw} was used to calculate these ratios,
and the result was cross-checked with the MSTW2008 PDF set~\cite{Martin:2009iq}.
The CTEQ and MSTW calculations agreed well and the uncertainty in this ratio does not
significantly contribute to the final result.
The most significant uncertainty in the limit computation is associated with our understanding
of the selection efficiency and detector acceptance ratio for $\cPZpr$ bosons
relative to the Z, denoted $R_\epsilon$.
The uncertainty in the total lepton selection efficiency at high mass
dominates the $R_\epsilon$
uncertainty.
The lepton selection efficiencies are measured in data up to
$\pt \sim 500$\GeV, but above 100\GeV the uncertainties in these measurements
become large.
This leads to a total uncertainty in $R_\epsilon$ of 3\% for the dimuon
channel and 8\% for the dielectron channel after including PDF uncertainties
in the acceptance.
The effects of misalignment, higher order corrections to the background shape,
and the uncertainty in backgrounds due to jets misidentified as leptons have
only
negligible
impact on the limits.
The upper limits on the ratio $R_\sigma$ for spin-1 and spin-2 particles obtained from the dilepton combined mass spectra are shown in Fig.~\ref{fig:ichepLimits}.
Table~\ref{tab:massLimits} shows the limits on $R_\sigma$ converted into mass limits on specific models.
The resonance is assumed to be narrow, meaning that the detector resolution dominates the width of the peak.
The $\ZPPSI$ with a relative width of 0.6\% is therefore considered narrow.
A wider resonance, such as the $\ZPSSM$, which has a width of 3\%, will have more background under the peak.
Consequently, we would set weaker limits on its production cross sections.
The two cases provide similar results when there is very little background after all selection criteria have been imposed.
This occurs around 1.4\TeV.
For a resonance below 1.4\TeV not to have been discovered, it must have a small coupling and therefore be narrow.
For the spin-2 case an additional requirement is that the ratio of $\Pg\Pg$ to $\cPq\cPaq$ production of the resonance
must be the same as the ratio for an RS graviton.
The combination of the 7 and 8\TeV data sets relies on this assumption, as $\Pg\Pg$ and
$\Pq\Paq$ cross sections scale differently with $\sqrt{s}$. For the spin-1 case,
no $\Pg\Pg$ coupling is considered.
The $\cPZpr$ and RS Graviton cross sections are calculated using the \PYTHIA event generator
with the CTEQ6.1 PDF set.
The LO cross sections are corrected for next-to leading (NLO) or NNLO QCD contributions using the same k-factors as Ref.~\cite{CMS-dilep-2011}.
 A mass dependent NNLO k-factor calculated
with \textsc{zwprodp}~\cite{Accomando:2010fz,Hamberg:1990np,vanNeerven:1991gh}
is used for the $\cPZpr$ models.
A flat NLO k-factor of 1.6 is applied to the RS graviton cross sections~\cite{Mathews:2005bw}.

\begin{figure*}[thb]
\begin{center}
\includegraphics[width=0.49\textwidth]{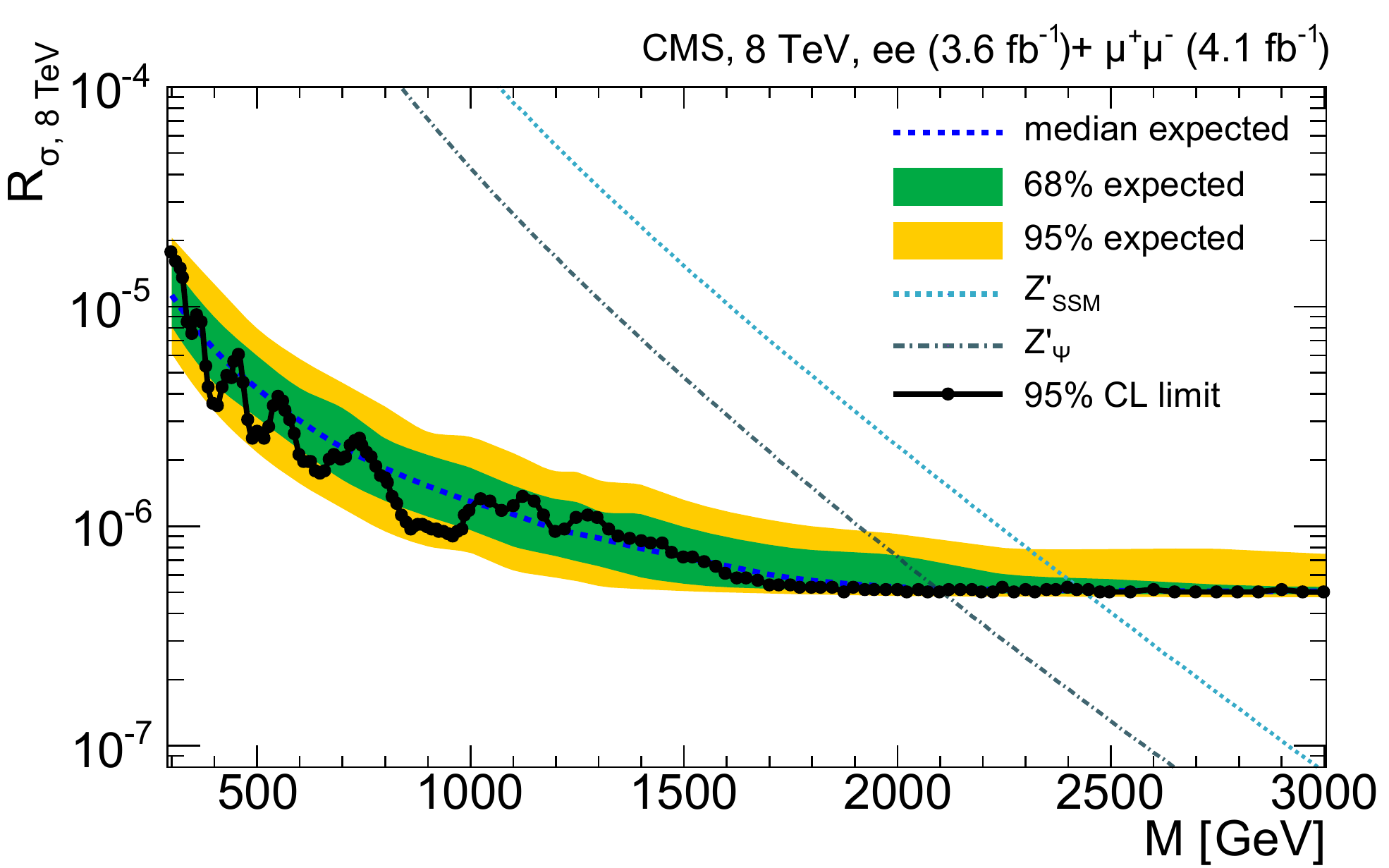}
\includegraphics[width=0.49\textwidth]{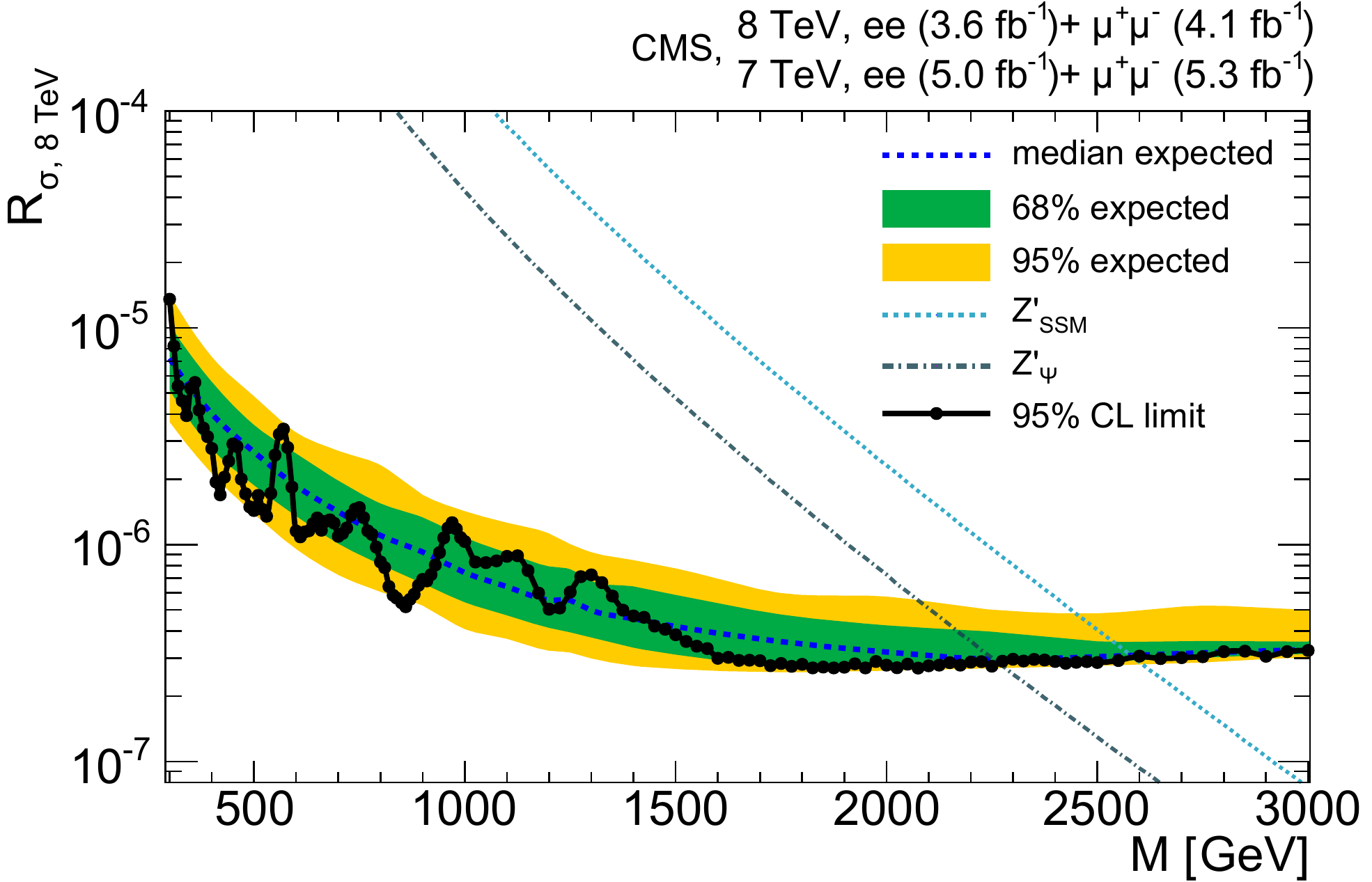}\\
\includegraphics[width=0.49\textwidth]{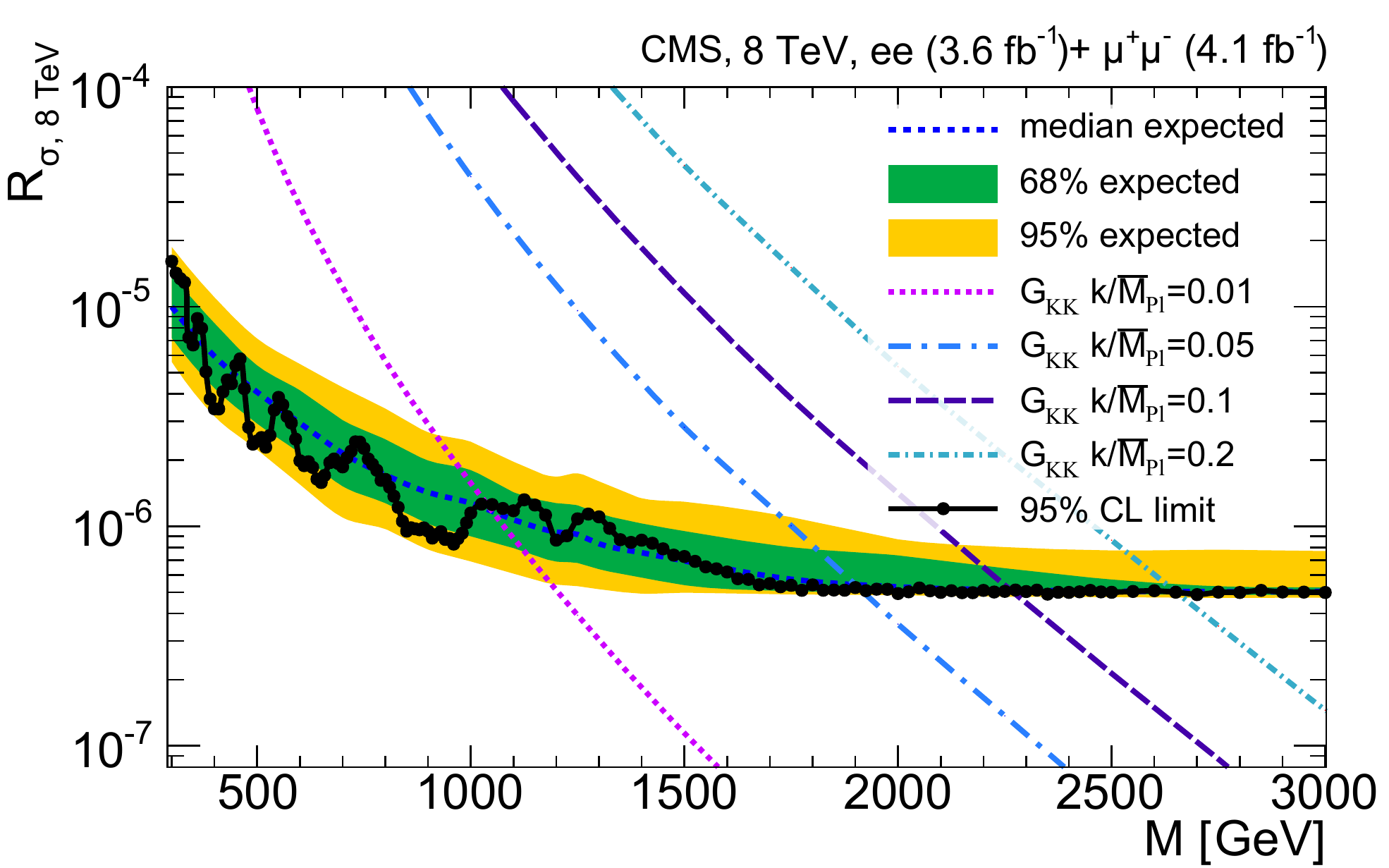}
\includegraphics[width=0.49\textwidth]{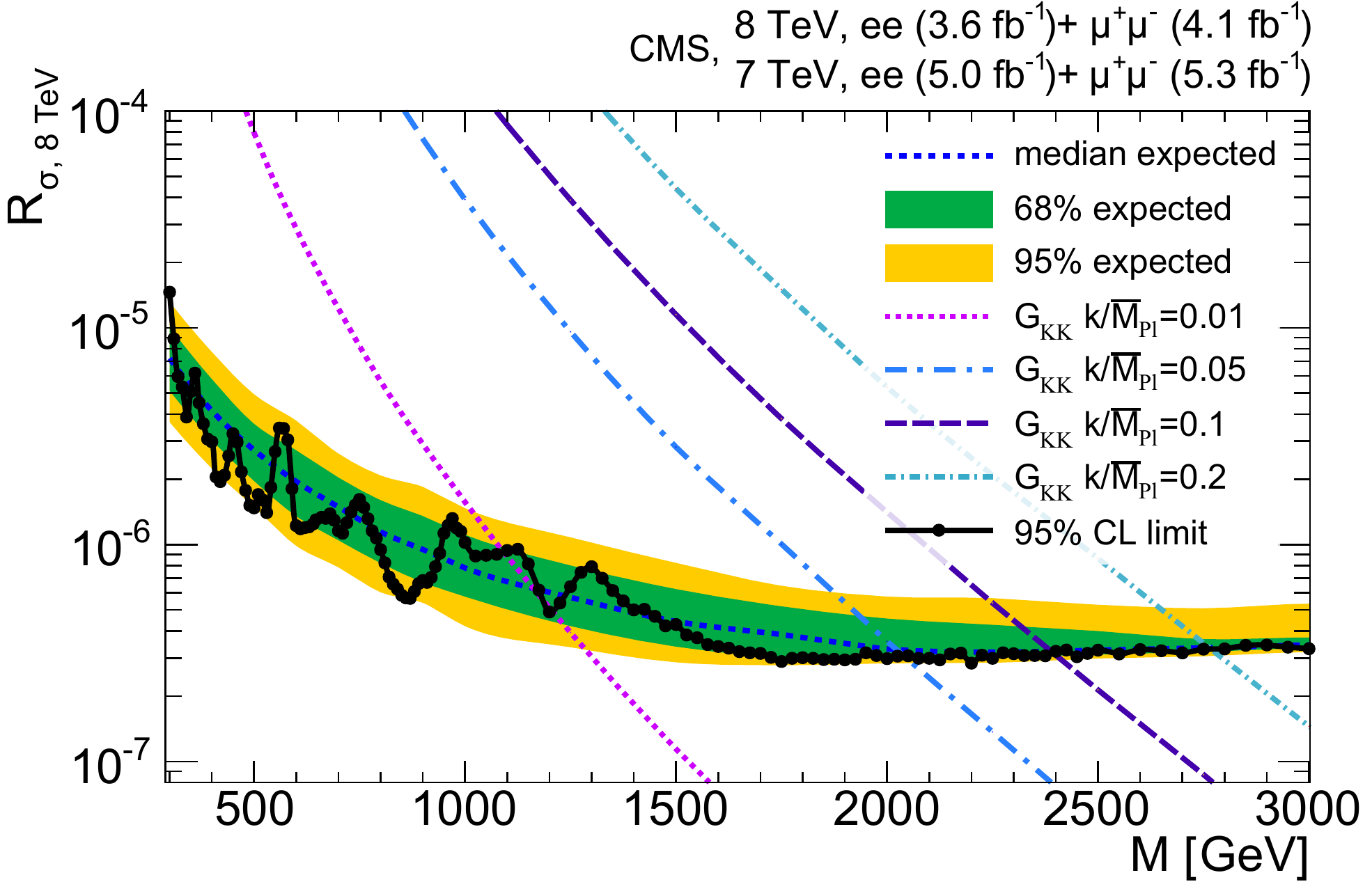}
\end{center}
\caption{
Upper limits on the ratio $R_{\sigma}$ of the production cross section times branching
fraction into lepton pairs to the same quantity for $\Zzero$ bosons, as a function of
resonance mass $M$ for spin-1 (top) and spin-2 (bottom) boson production.
The left plots are for the 8\TeV data set while the right plots are for the combination of the 7 and 8\TeV data sets.
For the spin-2 case, the 7 and 8\TeV data set combination
is only valid  for models that have the same fraction of $\Pq\Paq$
to $\Pg\Pg$ coupling as an RS graviton.
For the spin-1 case no $\Pg\Pg$ coupling is considered.
Shaded bands identified in the legend correspond to the 68\% and 95\% quantiles
for the expected limits, respectively.}
\label{fig:ichepLimits}
\end{figure*}

\begin{table}[htbp]
\begin{center}
\topcaption{Mass lower limits at the 95\% CL on specific models obtained using dilepton data at $\sqrt{s}=7$ and 8\TeV separately and combined. The 7\TeV results are taken from Ref.~\cite{CMS-dilep-2011}.}
\begin{tabular}{l|c|c|c} \hline\hline
\multirow{2}{*}{Model} & \multicolumn{3}{c}{Mass Limits (\GeVns{})} \\
 & 7\TeV & 8\TeV & 7+8\TeV \\\hline
$\ZPSSM$ & 2330 & 2440 & 2590 \\
$\ZPPSI$ &  2000 & 2110 & 2270 \\
$G_\mathrm{KK}$ ($k/\MPlanck = 0.1$) & 2140 & 2260  & 2390  \\
$G_\mathrm{KK}$ ($k/\MPlanck = 0.05$) & 1810  & 1900 & 2030 \\\hline\hline
\end{tabular}

\label{tab:massLimits}
\end{center}
\end{table}
\section{Summary}

The CMS Collaboration has searched for heavy narrow resonances in dimuon and dielectron
invariant mass spectra.
The search combined data samples from pp collisions at $\sqrt{s} = 7$\TeV~\cite{CMS-dilep-2011}
and $8$\TeV.
The $\sqrt{s} = 8$\TeV data sets have integrated luminosities of 4.1\fbinv (3.6\fbinv)
for the dimuon (dielectron) channel.
The $\sqrt{s} = 7$\TeV data sets have integrated luminosities of 5.3\fbinv (5.0\fbinv)
for the dimuon (dielectron) channel,
and have been previously published \cite{CMS-dilep-2011}.
The measured dilepton mass spectra are consistent with predictions from the standard model.
Upper limits on the cross section times branching fraction for the production
of new heavy narrow resonances relative to $\Zzero$ boson production are presented.
The findings exclude, at 95\% CL, a $\ZPSSM$ with standard model-like couplings below 2590\GeV\
and the superstring-inspired $\ZPPSI$ below 2260\GeV.
An RS graviton with $k/\MPlanck$ of 0.1 (0.05) is excluded below 2390 (2030)\GeV.
These are the most restrictive limits to date for the classes of models considered.
\section*{Acknowledgements}

We congratulate our colleagues in the CERN accelerator departments for the excellent performance of the LHC and thank the technical and administrative staffs at CERN and at other CMS institutes for their contributions to the success of the CMS effort. In addition, we gratefully acknowledge the computing centres and personnel of the Worldwide LHC Computing Grid for delivering so effectively the computing infrastructure essential to our analyses. Finally, we acknowledge the enduring support for the construction and operation of the LHC and the CMS detector provided by the following funding agencies: BMWF and FWF (Austria); FNRS and FWO (Belgium); CNPq, CAPES, FAPERJ, and FAPESP (Brazil); MEYS (Bulgaria); CERN; CAS, MoST, and NSFC (China); COLCIENCIAS (Colombia); MSES (Croatia); RPF (Cyprus); MoER, SF0690030s09 and ERDF (Estonia); Academy of Finland, MEC, and HIP (Finland); CEA and CNRS/IN2P3 (France); BMBF, DFG, and HGF (Germany); GSRT (Greece); OTKA and NKTH (Hungary); DAE and DST (India); IPM (Iran); SFI (Ireland); INFN (Italy); NRF and WCU (Republic of Korea); LAS (Lithuania); CINVESTAV, CONACYT, SEP, and UASLP-FAI (Mexico); MSI (New Zealand); PAEC (Pakistan); MSHE and NSC (Poland); FCT (Portugal); JINR (Armenia, Belarus, Georgia, Ukraine, Uzbekistan); MON, RosAtom, RAS and RFBR (Russia); MSTD (Serbia); SEIDI and CPAN (Spain); Swiss Funding Agencies (Switzerland); NSC (Taipei); ThEPCenter, IPST and NSTDA (Thailand); TUBITAK and TAEK (Turkey); NASU (Ukraine); STFC (United Kingdom); DOE and NSF (USA).

\bibliography{auto_generated}
\cleardoublepage \appendix\section{The CMS Collaboration \label{app:collab}}\begin{sloppypar}\hyphenpenalty=5000\widowpenalty=500\clubpenalty=5000\textbf{Yerevan Physics Institute,  Yerevan,  Armenia}\\*[0pt]
S.~Chatrchyan, V.~Khachatryan, A.M.~Sirunyan, A.~Tumasyan
\vskip\cmsinstskip
\textbf{Institut f\"{u}r Hochenergiephysik der OeAW,  Wien,  Austria}\\*[0pt]
W.~Adam, E.~Aguilo, T.~Bergauer, M.~Dragicevic, J.~Er\"{o}, C.~Fabjan\cmsAuthorMark{1}, M.~Friedl, R.~Fr\"{u}hwirth\cmsAuthorMark{1}, V.M.~Ghete, N.~H\"{o}rmann, J.~Hrubec, M.~Jeitler\cmsAuthorMark{1}, W.~Kiesenhofer, V.~Kn\"{u}nz, M.~Krammer\cmsAuthorMark{1}, I.~Kr\"{a}tschmer, D.~Liko, I.~Mikulec, M.~Pernicka$^{\textrm{\dag}}$, D.~Rabady\cmsAuthorMark{2}, B.~Rahbaran, C.~Rohringer, H.~Rohringer, R.~Sch\"{o}fbeck, J.~Strauss, A.~Taurok, W.~Waltenberger, C.-E.~Wulz\cmsAuthorMark{1}
% ...
\vskip\cmsinstskip
\dag:~Deceased\\
1:~~Also at Vienna University of Technology, Vienna, Austria\\
2:~~Also at CERN, European Organization for Nuclear Research, Geneva, Switzerland\\
% ...

\end{sloppypar}
\end{document}